\documentclass{webofc}

\usepackage[varg]{txfonts}   
\usepackage{hyperref}
\usepackage{url}
\hypersetup{colorlinks=true,citecolor=blue,urlcolor=blue,linkcolor=blue}
\begin{document}
\title{\boldmath Search for Light $H^+$ via $e^+e^- \to H^+H^- \to c\bar b\bar cb$ @ ILC500}
%
%

\author{\firstname{George Wei-Shu} \lastname{Hou}$^1$\fnsep\thanks{wshou@phys.ntu.edu.tw} 
}

\institute{Department of Physics, National Taiwan University, Taipei 10617, Taiwan}

\abstract{Our current {\it impasse}: no new physics seen beyond the $h(125)$ boson. We advocate 
the {\it general} two Higgs doublet model without $Z_2$ symmetry, allowing extra Yukawa 
couplings, where we present 5 clear merits. We report on the search effort at LHC for $pp \to tt\bar
 c$, then on the new effort of $pp \to bH^+ \to bt\bar b$ search for $H^+$, and comment on our 
\underline{\bf Decadal Mission} program in Taiwan. Finally, we turn to the $e^+e^- \to H^+H^-
 \to c\bar b\bar cb$ program at ILC500 as a {\it what-if} situation at the LHC, demonstrating light 
$H^+$ reconstruction without $c$-tagging. We then end with discussion and conclusion.}
\vskip-0.5cm
\maketitle
\section{Introduction: Our current {\it impasse}}
\label{intro}

\hskip0.525cm Our current {\it impasse}: ${\cal N}$o ${\cal N}$ew ${\cal P}$hysics (${\cal NNP}$) beyond the {\it SM}-like $h(125)$ boson.

The apt and observant science reporter, Adrian Cho, wrote a warning on March 23, 2007 in {\it Science}: {\bf the Nightmare Scenario -- The Higgs and Nothing Else}, quoting Jon Ellis that 
``it would be the real five star disaster, because that would mean there wouldn't need to be any new 
physics.'' Just before the ten-year anniversary celebration of the {\it landmark} Higgs discovery at 
CERN, Adrian Cho wrote a reminder in June: Ten years after the Higgs, physicists face the 
Nightmare of finding nothing else, and stressed that: ``Unless Europe's Large Hadron 
Collider {\it coughs} up a surprise, the field of particle physics may {\it wheeze} to its end.'' 
Such seems the plight of particle physics. Or maybe not, as we shall explore.

We dare think differently. What we tote in Sec.\;2 is the {\it general} 2 Higgs Doublet Model 
(g2HDM) as the\,{\it Next\,New\,Physics}\,({\it NNP}): by dropping the usual $Z_2$ symmetry, 
there are two sets of new dim.-4 couplings, viz.\,extra Yukawa couplings that are fit for electroweak 
baryogenesis (EWBG), with ${\cal O}(1)$ quartic couplings providing the 1$^{\rm st}$ order 
EW Phase Transtion (1$^{\rm st}$oEWPT). The take-home message: Don't EFT {\it yet!} We 
give 5 clear merits of g2HDM. In Sec.\;3, we present our \underline{\bf Decadal Mission of the 
New Higgs/Flavor Era}, presenting a midterm report of a flagship project in Taiwan: my view 
on {\it beyond SM} ({\it BSM}) physics, that ATLAS and CMS published their g2HDM-motivated 
search for $pp \to tH/tA \to tt\bar c$. In Sec.~4, we present a new $pp \to bH^+ \to bt\bar b$ and $pp \to tt\bar t$ search program. In Sec.~5, we turn to the title subject of $e^+e^- \to H^+H^- \to
 c\bar b\bar cb$ @ ILC500, as a special {\it what-if} situation at LHC. 
After some discussion, we offer our conclusion: from ${\cal NNP}$ to {\it NNP}.

\section{General Two Higgs Doublet Model}
\label{g2HDM}

The discovery of the 125 GeV Higgs boson between ATLAS~\cite{ATLAS:2012yve} and CMS~\cite{CMS:2012qbp} was a true {\it landmark}, as we have never seen a fundamental scalar before. But having discovered one doublet, with no theorem forbidding another, 2HDM should be 
a no-brainer. We, however, advocate g2HDM by dropping the {\it ad hoc} assumption of a $Z_2$ symmetry, in accord with my maxim that any additional assumption would cost an 
${\cal O}(\alpha)$ in likelihood to realize.

Let us now give the five {\bf Merit}s of g2HDM.
{\bf Merti-1}: extra top Yukawa couplings $\rho_{tt}, \rho_{tc}$ at ${\cal O}(1)$ and naturally 
complex, can {\it each}~\cite{Fuyuto:2017ewj} drive EWBG; the leading and robust driver is
$\lambda_t\,{\rm Im}\,\rho_{tt}$. Interestingly, ${\cal O}(1)$ Higgs quartic couplings of g2HDM, 
in all 7 of them, can provide 1$^{\rm st}$oEWPT~\cite{Kanemura:2004ch}, which implies 
primordial gravitational waves that are detectable by space-based probes. But with ${\cal O}(1)$ 
CP Violation (CPV) fit for baryogenesis, one is highly vulnarable to electron EDM (eEDM) 
constraints, such as ACME~\cite{ACME:2018yjb} and JILA~\cite{Roussy:2022cmp}, the cutting 
edge of CPV search teeming with competitive AMO experiments. How to evade these?

\begin{figure}[h]
\centering
\includegraphics[width=6cm]{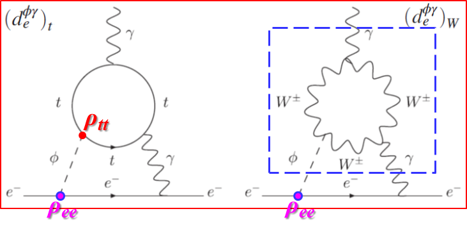}
\caption{With $\phi = h$, $H$, $A$ (even $H^+$), the $\phi$-$\gamma$-$\gamma^*$ loop 
insertion gives the main eEDM diagram.}
\label{fig-1}       
\end{figure}

{\bf Merit-2} of g2HDM is the discovery of a spectacular 2-loop diagrammatic cancellation mechanism~\cite{Fuyuto:2019svr} between top and $W$-loops of Fig.~\ref{fig-1}, {\it without} 
invoking SUSY. The upshot is 
\begin{align}
   |\rho_{ee}/\rho_{tt}| \sim \lambda_e/\lambda_t,
\label{flav-enigma}
\end{align}
with the prerequisite condition of a phase-lock: $\arg\rho_{ee} = - \arg\rho_{tt}$. See 
Ref.~\cite{Fuyuto:2019svr} for more discussion.
Eq.~(\ref{flav-enigma}) seems to imply that the $\ell$-type and $u$-type Yukawa matrices 
{\it know} the mass and mixing hierarchies that we learned from the SM sector. As we will 
elaborate later, this suggests that we may have broken ``the flavor code'', i.e. why {\it Nature} 
implemented the flavor-engima, the mass and mixing hierarchies in the first place.

{\bf Merit-3} of g2HDM concerns Glashow's worry about Flavor Changing Neutral Couplings~\cite{Glashow:1976nt} (FCNC), such as $t \to ch$. But once we discovered $h(125)$ is 
lighter than the top, {\bf our PDG duty}~\cite{ParticleDataGroup:2022pth} is to search for it! The 
$t \to ch$ decay was first pointed out in Ref.~\cite{Hou:1991un} as a thing to watch in 2HDM 
without $Z_2$ (originally called 2HDM-III, i.e. neither 2HDM-I nor 2HDM-II), and stressed that it 
is naturally ``flavor-controlled'', i.e. suppressed by mass and mixing hierarchies. In a follow-up 
paper~\cite{Chen:2013qta} after $h(125)$ discovery, it was pointed out that the coupling should be 
$\rho_{tc}\cos\gamma$, where $c_\gamma \equiv \cos\gamma$ is the $h$-$H$ mixing angle, with 
$H$ the exotic CP-even scalar. Curiously, $t \to ch$ remains elusive to 
date~\cite{ParticleDataGroup:2022pth}. Perhaps someone should tell Glashow: Fear not! 
{\it Nature} threw in {\it alignment} -- small $h$-$H$ mixing, i.e. small $c_\gamma$. As
{\it alignment} is quite {\it emergent} a phenomena that no one could have predicted. Who would 
have guessed that {\it Nature} would throw in something unrelated to {\it flavor} to control FCNC!? 

{\bf Merit-4} of g2HDM: small $c_\gamma$ does {\it not} contradict ${\cal O}(1)$ quartics~\cite{Hou:2017hiw}. Turning the argument around, one can argue that the exotic scalars 
$H$, $A$ and $H^+$ populate 300--600~GeV.

{\bf Merit-5} of g2HDM: with $t \to ch$ suppressed by {\it alignment} and with $H$, $A$ and 
$H^+$ populating 300--600~GeV, it is then natural to pursue 
\begin{align}
cg \to tH/tA \to tt\bar c,
\label{ttc}
\end{align}
production at the LHC~\cite{Kohda:2017fkn}. It was subsequently found that a better process 
is~\cite{Ghosh:2019exx} 
\begin{align}
cg \to bH^+ \to bt\bar b, 
\label{btb}
\end{align}
where one has a recoiling $b$ rather than $t$. This lowers the $H^+$ threshold and allows a more 
complete range of $m_{H^+}$ to be probed.

\section{\boldmath Midterm Report of Decadal Mission: $pp \to tt\bar c$ at ATLAS \& CMS}
\label{decadal}

\begin{figure}[h]
\centering
\includegraphics[width=5cm]{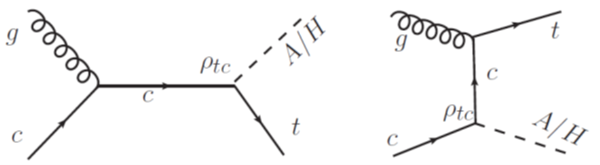}
\caption{Diagram for $cg \to tH/tA \to tt\bar c$, Eq.~(\ref{ttc}), with $H/A \to t\bar c$,
hence controlled by $s_\gamma \equiv \sin\gamma \to 1$.}
\label{fig-2}       
\end{figure}

As already noted, we hereby report that both ATLAS~\cite{ATLAS:2023tlp} and 
CMS~\cite{CMS:2023xpx} have followed the suggestion of Ref.~\cite{Kohda:2017fkn}, and 
published their search results of Eq.~(\ref{ttc}), with the signature of same-sign top pair plus 
$c$-jet, which serves as a discriminator to suppress background, as depicted in Fig.~\ref{fig-2}. 
We note that $H/A \to t\bar t$ is also possible, though with higher threshold.

\vskip-0.3cm
We note that neither ATLAS~\cite{ATLAS:2023tlp} nor CMS~\cite{CMS:2023xpx} saw any 
signal, which may not be too surprising. We therefore do not comment on this any further, except 
to outline the \underline{\bf Decadal Mission \it of the New Higgs/Flavor Era}, a flagship project 
of Taiwan, composed of 4 directions, each led by a Co-PI: 1) CMS analysis; 2) Belle~II analysis; 
3) lattice study of 1$^{\rm st}$oEWPT; and 4) steering by pheno studies (since 2017).

We are extremely thankful to the approval of the {\bf Decadal Mission} project by Taiwan's NSTC,
which allowed the timely build-up of the CERN-based CMS analysis team, such that the CMS
study only lagged ATLAS by 4 months.

\section{\boldmath Post-Midterm: $pp \to bH^+ \to bt\bar b$ and $pp \to tt\bar t$ at LHC}
\label{PostMidterm}

The CMS team is now focused on two new directions: Eq.~(\ref{btb}), as well as Eq.~(\ref{ttc}) 
but with $H/A \to t\bar t$, namely triple-top final state, which is highly suppressed in SM (ATLAS 
study~\cite{ATLAS:2023tlp} had covered it).
As LHC Run~3 is now progressing well, when data reaches Run~2 level, the CMS team would 
redo $t \to ch$ and $pp \to tt\bar c$. All four processes have discovery potential, which follows 
the plan according to our {\bf Decadal Mission} narrative~\cite{Hou:2021wjj}.

For Belle~II, the main target is to measure ${\cal B}(B \to \mu\nu)/{\cal B}(B \to \tau\nu)$. It was 
first pointed out~\cite{Chang:2017wpl} in an experimental FPCP review that this ratio provides an 
interesting test of g2HDM, and later demonstrated by explicit~\cite{Hou:2019uxa} calculation. For 
more discussion of flavor probes in g2HDM, see Ref.~\cite{Hou:2020itz}, especially the rich 
picture-table of Fig.~3 there.

\section{\boldmath $e^+e^- \to H^+H^- \to c\bar b\bar cb$ @ ILC500}
\label{ILC500}

\begin{figure}[h]
\centering
\includegraphics[width=10.5cm]{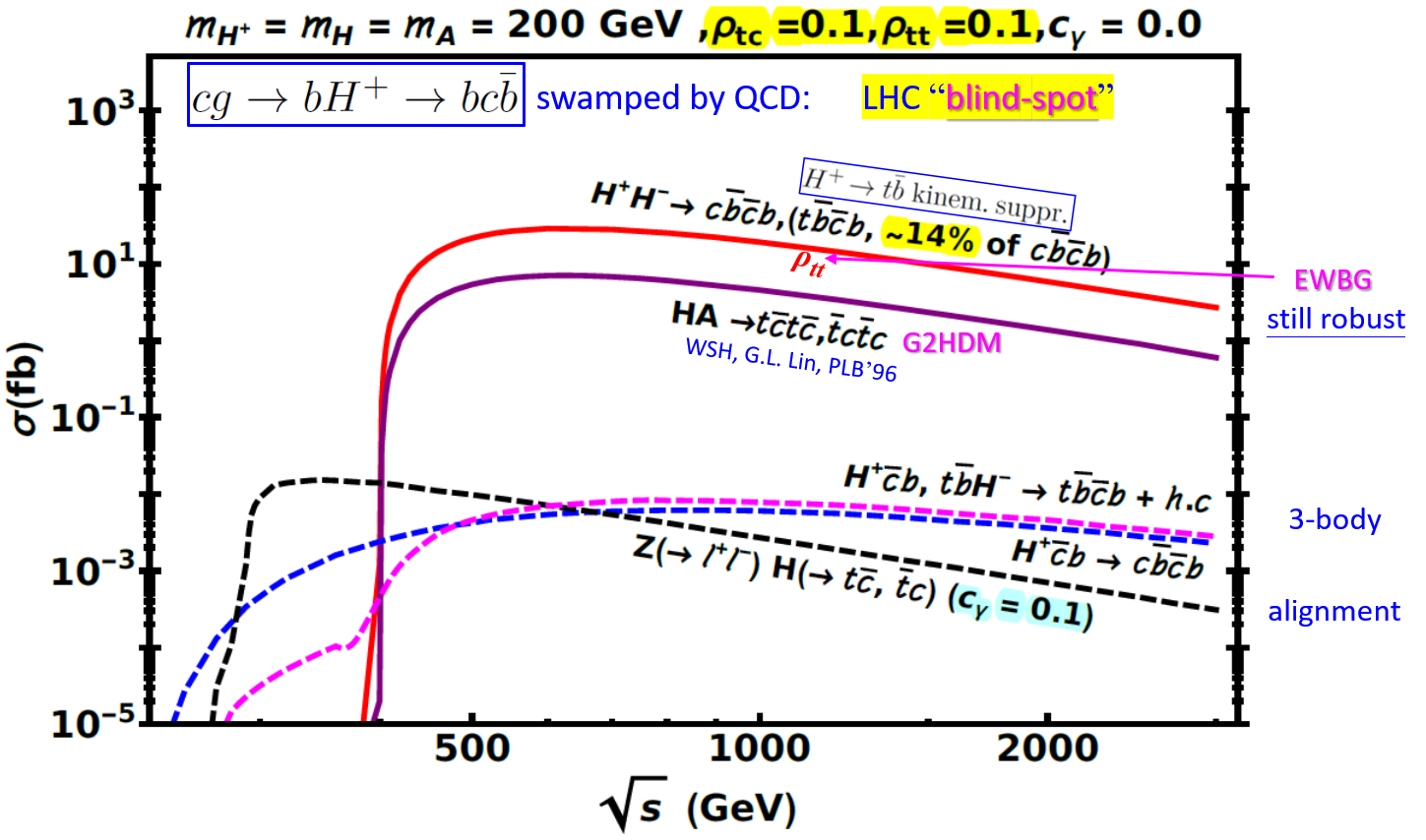}
\caption{``Blind-spot'' scenario for nondiscovery of $H/A$ or $H^+$ at LHC, though EWBG is 
still robust.}
\label{fig-3}       
\end{figure}

Finally, we turn to the subject title of our talk: light $H^+$ search at ILC500.
Fig.~\ref{fig-3} depicts a {\it what-if} situation~\cite{Hou:2021qff} at the LHC: 
{\it nonobservation!?} This could happen more subtly, such as $\rho_{tc} \simeq \rho_{tt}
 \simeq 0.1$ suppressing production. Furthermore, suppose $m_{H^+} = m_H = m_A =$ 
200~GeV, then $H$ and $A$ would mutually cancel\;\cite{Kohda:2017fkn}, while $cg \to bH^+
 \to bc\bar b$ would be buried in QCD background. This would constitute an LHC ``blind-spot'', 
though EWBG is still robust~\cite{Fuyuto:2017ewj}!

This ``blind-spot'' can be unraveled at ILC500. Note $H^+ \to t\bar b$ is suppressed to 14\%. 
The $e^+e^- \to HA \to t\bar ct\bar c$ process\,\cite{Hou:1995qh}\,is not far below, but the 
same-sign top pair signature to suppress background means $H$ and $A$ reconstruction would 
not be easy, so at best it serves as confirmation. The 3-body $H^+\bar cb$ process is kinematically 
suppressed, while the $Z(\ell^+\ell^-)H(tc)$ tag mode is suppressed by {\it alignment}, falling 
below 3-body. So, how to proceed?

\begin{figure}[h]
\centering
\includegraphics[width=8.75cm]{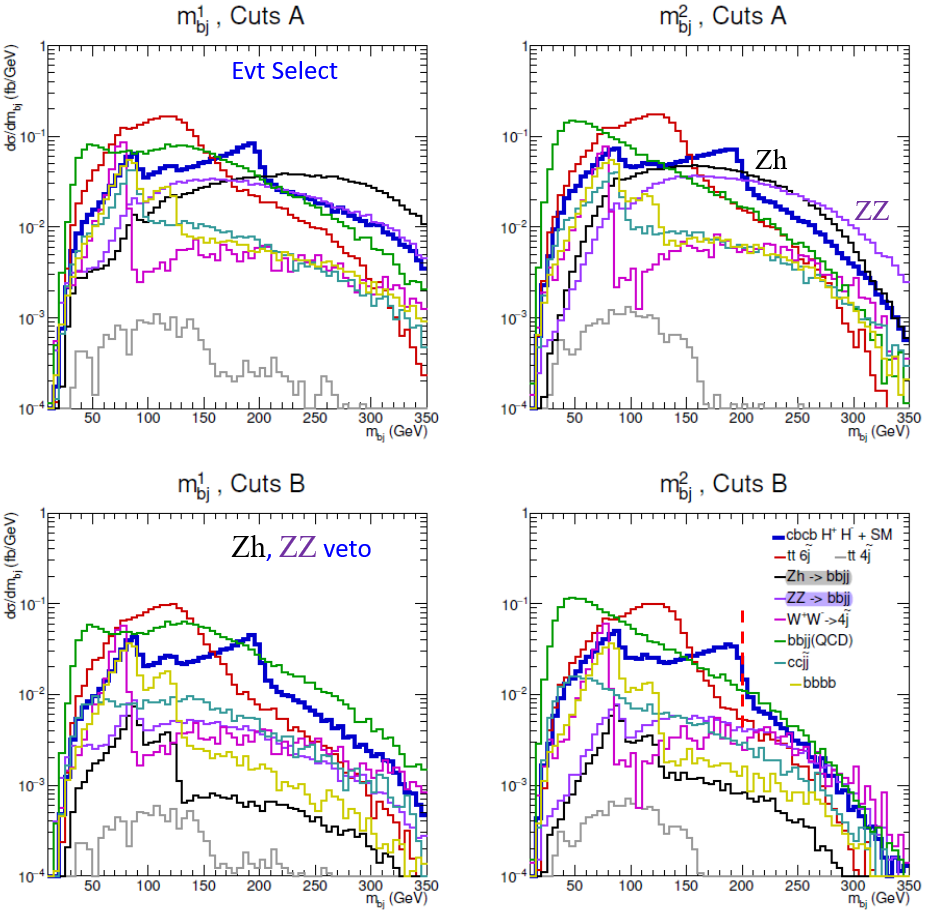}
\caption{Cut A, where the light ($c$) jet closest in angular proximity is paired with the $b$-jet,
forming $m_{bj}^1$ and $m_{bj}^2$. Cut B then vetos $Z, h \to b\bar b$, where the drop is
visible.}
\label{fig-4}       
\end{figure}

\vskip-0.3cm
We generate events by~\cite{Hou:2021qff} running Madgraph + Feyrules + PYTHIA6.4 + 
Delphes3.5.0 with ILD card. Treating the $c$-jet as a light jet, we select events by pairing the light 
jet by angular proximity to the $b$-jet. We call this selection Cut A, with two pairings $m_{bj}^1$ 
and $m_{bj}^2$. Since $Zh$ and $ZZ$ (together with $t\bar t$) are the main backgrounds, while 
both $h$ and $Z$ are rich in $b\bar b$, we veto $Z, h \to b\bar b$, which we call Cut B, where the 
drop is visible in Fig.~\ref{fig-4}.

We calculate significance via
\begin{equation}
    Z(n|n_{\rm pred}) = \sqrt{-2\,ln\frac{L(n|n_{\rm pred})}{L(n|n)}},
\label{eq:Cowan}
\end{equation}
where $L(n_1|n_0) = e^{-n_1}n_1^{n_0}/n_{0}!$. With Cuts A and B, the significance is already 
above 20, but the real issue is reconstructing the $H^+$ mass, where we devise Cut C:
\begin{equation}
|m_{bj}^{1} - m_{bj}^{2}| < 0.1\times m_{bj}^{1}.
\end{equation}
The 0.1 factor is somewhat arbitrarily chosen to retain statistics, with no attempt at optimization 
to make $m_{bj}^1 = m_{bj}^2$ more restrictive. Experiment can do much better, including
adding $c$-tagging (in answering a question from the audience, of course ILC has workable
$c$-tagging, which can only help in affirming the signal). 

\begin{figure}[h]
\centering
\includegraphics[width=10cm]{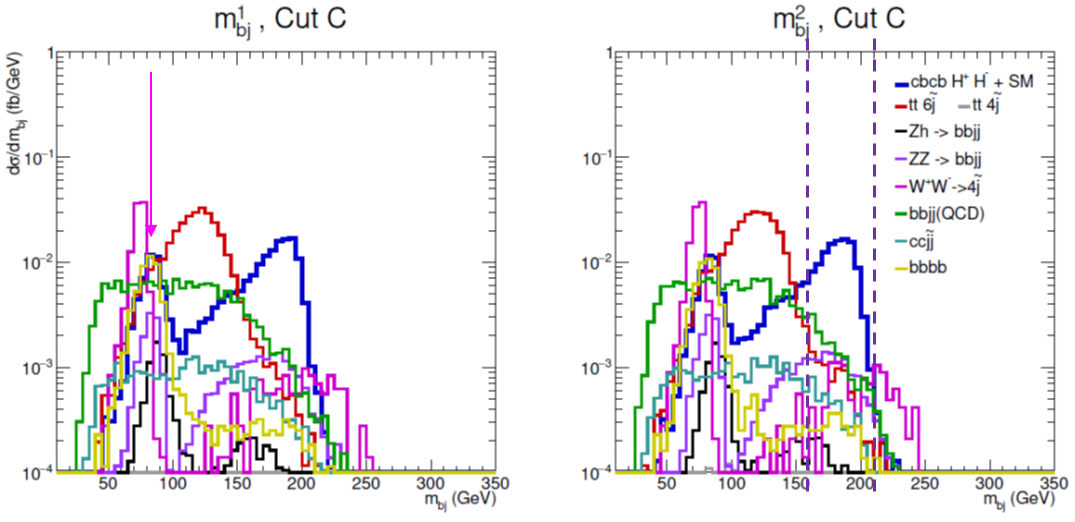}
\caption{Result of Cut C; the pink arrow on the left-side $m_{bj}^1$ plot points at the 
$W \to c\bar s$ ``edge''.}
\label{fig-5}       
\end{figure}
\vskip-0.3cm
Fig.~\ref{fig-5} shows the result with Cut C, which is quite revealing.~We had not vetoed 
the~$W$, and note that the pink arrow on left plot of Fig.~\ref{fig-5} points to the $W \to cs$ 
Jakobian-drop, where DELPHES tags the $c$-jet as $b$-jet. We then note that the blue signal of 
$c\bar b\bar cb$ also shows an ``edge'', or Jakobian behavior. It turns out that Cut~C does pair the 
correct $c$-jet with the $b$-jet --- without inputting any mass information, nor activating 
$c$-tagging but suppresses much background! The signal ``edge'' reveals the $H^\pm$ mass at 
200~GeV. The key to our success is a result of the two-body $H^+H^-$ production, followed by 
$H^\pm \to cb$ two-body decay.

With $m_{H^+} \simeq 200$~GeV revealed, one can do a ``mass cut'': integrating events in the 
mass window of
\begin{align}
160~{\rm GeV} < m_{bj}^{2} < 210~{\rm GeV},
\end{align}
with further reduction of background, and the significance reaches 26.3 (significance had dropped
to 11.1 with plain Cut C including the full range background).

\section{Discussion and Conclusion}
\label{disc}

Pairing by angular proximity improves~\cite{Hou:2021qff} with energy, e.g. at 1 TeV, the ``Edge'' 
is sharper. But at the $pp$ collider, the produced $H^+H^-$ system is not in the CM frame, and 
with $p_T$ of the system unknown, while cross sections are generally smaller than QCD processes.

To summarize our \underline{\bf Decadal Mission} effort~\cite{Hou:2021wjj}, it is still my goal 
and hope that we can discover the exotic $H$, $A$ and $H^+$ bosons around 500~GeV or so at 
the LHC, where one could account for baryogenesis while evading eEDM! That would be plainly 
fantastic, and with a large number of FPCP probes~\cite{Hou:2020itz} thrown in, which is exciting.

But what we have presented is a {\it perverse} but possible {\it what-if} situation: 
$\rho_{tc} = \rho_{tt} = 0.1$ with $H$, $A$ and $H^+$ degenerate at 200~GeV. What we have 
shown is that even such a case can still be covered at the 500~GeV ILC (or higher: CLIC?) through 
$e^+e^- \to H^+H^- \to c\bar b\bar cb$, where one can reconstruct $m_{H^+}$, even without 
invoking $c$-tagging (which can only help). In addition, one can probe $e^+e^- \to HA \to t\bar c
 t\bar c$ to affirm\cite{Hou:1995qh} g2HDM. The strength of $\rho_{tt} \simeq 0.1$ can be 
measured through the kinematically suppressed $e^+e^- \to H^+H^- \to t\bar b\bar cb$ mode, and 
affirm EWBG. Actual measurement to probe CPV would need further study.

To conclude: from $\cal NNP$ ($\cal N$o $\cal N$ew $\cal P$hysics) at LHC, to {\it NNP} 
({\it Next New Physics}) in g2HDM, {\it if} this is {\it Nature}'s chosen path.

\

\noindent{\bf Acknowledgement.} This work is supported by NSTC 112 2639-M-002-006-ASP
of Taiwan. I thank Rishabh and Tanmoy for their contribution to this fun effort.

%
%

\end{document}